\documentclass[12pt]{article}
\usepackage{amsmath,amssymb,amsfonts}
\usepackage[paper=letterpaper,margin=1.0in]{geometry}

\newcommand{\be}{\begin{equation}}
\newcommand{\ee}{\end{equation}}
\newcommand{\bea}{\begin{eqnarray}}
\newcommand{\eea}{\end{eqnarray}}
\newcommand{\ba}{\begin{array}}
\newcommand{\ea}{\end{array}}
\newcommand{\ben}{\begin{enumerate}}
\newcommand{\een}{\end{enumerate}}
\newcommand{\bi}{\begin{itemize}}
\newcommand{\ei}{\end{itemize}}
\newcommand{\bc}{\begin{center}}
\newcommand{\ec}{\end{center}}
\newcommand{\bt}{\begin{table}}
\newcommand{\et}{\end{table}}
\newcommand{\btab}{\begin{tabular}}
\newcommand{\etab}{\end{tabular}}
\newcommand{\bs}{\begin{slide}}
\newcommand{\es}{\end{slide}}

\begin{document}

\def\CA{{\cal A}}
\def\CB{{\cal B}}
\def\CC{{\cal C}}
\def\CD{{\cal D}}
\def\CE{{\cal E}}
\def\CF{{\cal F}}
\def\CG{{\cal G}}
\def\CH{{\cal H}}
\def\CI{{\cal I}}
\def\CJ{{\cal J}}
\def\CK{{\cal K}}
\def\CL{{\cal L}}
\def\CM{{\cal M}}
\def\CN{{\cal N}}
\def\CO{{\cal O}}
\def\CP{{\cal P}}
\def\CQ{{\cal Q}}
\def\CR{{\cal R}}
\def\CS{{\cal S}}
\def\CT{{\cal T}}
\def\CU{{\cal U}}
\def\CV{{\cal V}}
\def\CW{{\cal W}}
\def\CX{{\cal X}}
\def\CY{{\cal Y}}
\def\CZ{{\cal Z}}

\newcommand{\todo}[1]{{\em \small {#1}}\marginpar{$\Longleftarrow$}}
\newcommand{\labell}[1]{\label{#1}\qquad_{#1}} 
\newcommand{\bbibitem}[1]{\bibitem{#1}\marginpar{#1}}
\newcommand{\llabel}[1]{\label{#1}\marginpar{#1}}

\newcommand{\pa}{\partial}
\renewcommand{\d}[1]{{\rm d}{#1}}
\newcommand{\BC}{\mathbb{C}}
\newcommand{\BR}{\mathbb{R}}
\newcommand{\BZ}{\mathbb{Z}}
\newcommand{\li}{{\rm li}}
\newcommand{\Li}{{\rm Li}}
\newcommand{\bra}[1]{\langle{#1}|}
\newcommand{\ket}[1]{|{#1}\rangle}
\newcommand{\vev}[1]{\langle{#1}\rangle}
\newcommand{\R}{{\rm Re}}
\newcommand{\I}{{\rm Im}}

{\footnotesize
\rightline{IHES/P/08/35}
\rightline{VPI-IPNAS-08-12}
}

\centerline{\Large \bf }\vskip0.25cm
\centerline{\Large \bf }\vskip0.25cm
\centerline{\Large \bf Turbulence and Holography}
\vskip0.25cm
\vskip 1cm

\renewcommand{\thefootnote}{\fnsymbol{footnote}}
\centerline{{\bf
Vishnu Jejjala${}^{1}$\footnote{\tt vishnu@ihes.fr},
Djordje Minic${}^{2}$\footnote{\tt dminic@vt.edu},
Y.\ Jack Ng${}^{3}$\footnote{\tt yjng@physics.unc.edu}, and
Chia-Hsiung Tze${}^{2}$\footnote{\tt kahong@vt.edu}}}
\vskip .5cm
\centerline{${}^1$\it Institut des Hautes \'Etudes Scientifiques}
\centerline{\it 35, Route des Chartres}
\centerline{\it 91440 Bures-sur-Yvette, France}
\vskip .5cm
\centerline{${}^2$\it Institute for Particle, Nuclear and Astronomical Sciences}
\centerline{\it Department of Physics, Virginia Tech}
\centerline{\it Blacksburg, VA 24061, U.S.A.}
\vskip .5cm
\centerline{${}^3$\it Institute of Field Physics}
\centerline{\it Department of Physics and Astronomy, University of North Carolina}
\centerline{\it Chapel Hill, NC 27599, U.S.A.}
\vskip .5cm

\begin{abstract}
We examine the interplay between recent advances in quantum gravity and the problem of turbulence.
In particular, we argue that in the gravitational context the phenomenon of turbulence is intimately related to the properties of spacetime foam.
In this framework we discuss the relation of turbulence and holography and the interpretation of the Kolmogorov scaling in the quantum gravitational setting.
\end{abstract}

\setcounter{footnote}{0}
\renewcommand{\thefootnote}{\arabic{footnote}}

\newpage

\section{Introduction}

Turbulence stands as one of the towering unsolved problems of classical physics~\cite{landau, frisch, monin}.
This problem has so many facets that it seems almost overwhelming.
One of the most puzzling aspects and yet strikingly simple to state is the famous Kolmogorov scaling~\cite{k41}, which specifies the behavior of $n$-point correlation functions of the fluid velocity.
While in real fluids, this scale invariance is empirically broken, one may still reasonably expect it to be restored, in a statistical sense, in the limit of infinite Reynolds number.

Here we try to approach Kolmogorov scaling by employing some recent ideas from quantum gravity.
In view of the fact that the fundamental equations of turbulent fluids at very high Reynolds numbers are invariant under volume preserving diffeomorphisms, and given that at least on the na\"{\i}ve level, the fundamental symmetries of quantum gravity at low-energies are spacetime diffeomorphisms, perhaps the connection between turbulence and quantum gravity should not be completely surprising.

In fact several recent papers have explored various aspects of fluid dynamics from the perspective of quantum gravity~\cite{sonetal, jp, ns, shiraz, makoto}.\footnote{
As well, effective gravitational physics was recently argued to play an important r\^ole in other many-body problems~\cite{mbody}.}
In this note, as a first step, we similarly bridge two phenomenologies, that of incompressible fluids and that of spacetime foam~\cite{wheeler}.
Specifically we relate Kolmogorov's scaling of fully developed turbulence and a generic holographic model of spacetime foam~\cite{ng}.

Our presentation is as follows.
Section~2 begins with a very brief review of Kolmogorov's 1941 and 1962 theories, known as K41 and K62, respectively.
Section~3 motivates the connections between quantum gravity, fluctuations in fluids, and turbulence while
Section~4 sketches a Wheeler--DeWitt styled formulation of spacetime foam.
Section~5 provides a 
map via scaling laws between spacetime foam, holography, and Kolmogorov's universality.
In Section~6, we lay down some future directions suggested by this work and state our conclusions.

\section{The Kolmogorov scaling}

The Kolmogorov~1941 scaling~\cite{k41} (also independently derived by Heisenberg~\cite{wh} and Onsager~\cite{lo}) works in the infinite Reynolds number limit in which the viscosity term $\nu \nabla^2 \vec{v}$ in the Navier--Stokes equations can be neglected compared to the convective term~\cite{landau, frisch, monin}.
Thus the basic starting point is given by the Euler equation
\be
\frac{\partial \vec{v} }{\partial t} + (\vec{v} \cdot \nabla) \vec{v} = - \frac{\nabla p}{\rho} ~,
\ee
where $\nabla \cdot \vec{v} = 0$.\footnote{
Here we note that the Euler equation has an infinite dimensional geometric interpretation as it describes the geodesic flow on the group of volume preserving diffeomorphisms~\cite{mardsen}.
While the above equations are mostly applied in spatial dimensions two or three, they do hold in general dimension.}
Kolmogorov's observation is that in the presence of a constant energy flux
\be
\frac{v^2}{t} \sim \varepsilon ~,
\ee
there is a single length scale given by the velocity and time $\ell \sim v t$.
The scaling of velocity with $\ell$ is
\be
v \sim (\varepsilon \ell)^{1/3} ~.
\ee
Kolmogorov~1941 deduces that the statistical moments are
\be
\langle (\delta v(\ell))^n \rangle \sim C_n \varepsilon^{n/3} \ell^{n/3} ~,
\ee
where $\delta v(\ell) = v(r+\ell) - v(r)$ and again $\varepsilon$ is the mean energy dissipation rate per unit mass.
The $C_n$ are dimensionless, universal, and constant.

In particular, this implies that the two-point function of velocity goes as
\be
\langle v^i(\ell) v^j(0) \rangle \sim \ell^{2/3} \delta^{ij} ~.
\ee
This is the famous {\em two-thirds law}.
There are deviations from this behavior known as far back as Landau~\cite{landau} whose criticisms led to the response in K62~\cite{frisch, k62}.
There Kolmogorov dealt with the issue of anomalous values of the scaling exponents by taking into account the observed intermittency effects mainly due to vorticity.
Nevertheless the special cases of $n=2,3$ are most notable.
The $n=2$ case (two-thirds law) tells us that the energy spectrum follows the power law $k^{-5/3}$~\cite{landau, frisch}.
The $n=3$ case is one in which the coefficient is explicitly known and is universal.
Here, in the limit of infinite Reynolds number, we have the {\em four-fifths law}:
\be
\langle (\delta v(\ell))^3 \rangle = -\frac{4}{5} \varepsilon \ell ~.
\ee
The derivation assumes only the following: homogeneity, isotropy, and the finiteness of $\varepsilon$.
In particular, scale invariance is not invoked.
This scaling is noteworthy in being the only exact dynamical result, obtainable directly from the Navier--Stokes equations themselves~\cite{frisch}.

Hydrodynamics can be thought of as an effective field theory~\cite{forster} capturing the dynamics at large spatial and temporal scales.
Moreover the constant energy flux can be nicely interpreted in terms of a quantum field theoretic anomaly~\cite{polyakov}.
Various other quantum field theoretic aspects of fluid turbulence have been discussed in~\cite{bob,gaw}.
(See also~\cite{migdal}.)

The fundamental problem is a dynamical one, namely how to get from the deterministic to a statistical description.
Also, Kolmogorov's distribution is not the usual Gibbs distribution.
How does such a non-Gibbsian distribution leading to Kolmogorov's scalings emerge from the equations of fluid dynamics?
The dynamical questions concern the approach to scaling, namely how to explain the breakdown of Kolmogorov scaling from first principles such as from the Navier--Stokes equation,
{\em i.e.}, how to account for the anomalous values of the exponents of the many-point functions.
These deviations from universality point to the non-Gaussian, non-Gibbsian nature of the velocity distribution.
Furthermore, in contrast to the usual effective field theoretic study of long-time, long-distance behavior at scales much larger than the high-energy cutoff, in turbulence one is interested in the opposite regime, much shorter than the cutoff scale!
Thus the renormalization group (RG) analysis is peculiar:
instead of going from ultraviolet (UV) to infrared (IR), the scaling for turbulence goes in the reverse direction~\cite{gaw}.

In this paper we wish to point out that some of these basic puzzling features are very natural from the point of view of quantum gravity\footnote{
As we have mentioned in the introduction the connection between turbulence and quantum gravity is perhaps not surprising on the level of symmetries involved.
At least at low-energies, gravity is defined by diffeomorphism invariance.
On the other hand one of the defining symmetries in the problem of turbulence is volume preserving diffeomorphisms: the flow is incompressible~\cite{arnold}.
The volume preserving diffeomorphisms also naturally occur in quantum gravity in the treatment of Matrix theory~\cite{nambu}.}
thus offering a new perspective on the problem of turbulence.

\section{Quantum gravity and turbulence}

Why would gravity have anything to do with fluid dynamics?
In this section we recall the recent discussion of induced gravity in fluid dynamics~\cite{unruh, abh}.
In the case of irrotational fluids in three spatial dimensions an effective metric emerges, called the acoustic metric~\cite{unruh}.
(The viscous flow has been considered in~\cite{stonevisser}.)
This comes about by considering fluctuations of the fluid density $\rho$ and the velocity potential $\phi$ (the velocity $\vec{v} = \nabla \phi$).
The underlying spacetime action of the moving fluid is
\be
S = \int \d{{}^4x}\ [ \rho \dot{\phi} + \frac{1}{2} \rho (\nabla \phi)^2 + U(\rho)] ~,
\ee
where $U(\rho)$ is the effective potential that upon variation leads to equations of motion for $\rho$ and $\phi$ (the Euler continuity equation and the Bernoulli energy balance equation).
In other words, following the effective field treatment of~\cite{sch, stone} (and the nice summary in~\cite{abh}),
\be
\dot{\rho} + \nabla \cdot (\rho \vec{v}) = 0 ~,
\ee
and also
\be
\dot{\phi} + \frac{1}{2} {\vec{v}}^2 + \frac{\d{U}}{\d{\rho}} = 0 ~.
\ee
When these equations of motion are perturbed around the equilibrium values $\rho_0$ and $\phi_0$,
\be
\rho = \rho_0 + \rho' ~, \qquad \phi = \phi_0 + \varphi ~,
\ee
one is led to the equations for the fluctuations of the velocity potential $\varphi$ (after eliminating $\rho'$ because it occurs quadratically in the perturbed effective action)
\be
\left(\frac{\partial}{\partial t} + \nabla \cdot \vec{v}\right) \frac{\rho_0}{c^2} \left(\frac{\partial}{\partial t} + \nabla \cdot \vec{v}\right) = \nabla (\rho_0 \nabla \varphi) ~.
\ee
In particular, the equation for the fluctuations of the velocity potential can be written in a geometric form~\cite{unruh} of a harmonic Laplace--Beltrami equation:
\be
\frac{1}{\sqrt{-g}} \partial_a( \sqrt{-g} g^{ab} \partial_b \varphi) = 0 ~.
\ee
Here, apart from a conformal factor, the effective space time metric has the canonical ADM form~\cite{unruh, abh}
\be
\d{s}^2 = \frac{\rho_0}{c} [ c^2 \d{t}^2 - \delta_{ij}(\d{x^i} - v^i \d{t})(\d{x^j} - v^j \d{t})] ~,
\ee
where $c$ is the sound velocity and $v^i$ are the components of the fluid's velocity vector.
This is the fundamental observation:
because of dragging of the sound in a moving fluid, the spherical shell associated with a given emitted sound pulse shifts by $v\, \d{t}$ in a unit time interval, so that its location can be found by solving the equation
\be
(\d{\vec{r}} - \vec{v}\, \d{t})^2 = c^2 \d{t}^2
\ee
which effectively can be arranged to the above acoustic metric~\cite{unruh, abh}.
The sound then propagates along null curves defined by the acoustic metric.

We observe that in the above expression for the metric the velocity of the fluid $v^i$ plays the r\^ole of the shift vector $N^i$ which is the Lagrange multiplier for the spatial diffeomorphism constraint (the momentum constraint) in the canonical Dirac/ADM treatment of Einstein gravity.
A fluctuation of $v^i$ would imply, given the intuition of Kolmogorov and this dictionary between fluids and gravity, a fluctuation of the shift vector.
This is possible provided the metric of spacetime fluctuates, which is a very loose, intuitive, semiclassical definition of the spacetime foam.

Now, whence comes the effective gravitational dynamics?
One idea discussed by Visser in~\cite{abh} is the idea of induced gravity.
After integrating out the fluctuations of the velocity potential (viewed as as a scalar field in the gravitational metric) around a background that does not satisfy the fluid equations of motion, one can then obtain an effective action which is of the induced gravity type and which includes the Einstein--Hilbert term and the cosmological term as well as higher order terms.

More explicitly (for a review see~\cite{essay}) the above equation for the sound wave fluctuations comes from the effective action
\be
S_{\varphi, g_{ab}} = \int \d{{}^4x}\ \sqrt{-g}\ (g^{ab} \partial_a \varphi \partial_b \varphi) ~.
\ee
By expanding $\varphi$ around a fixed configuration and by integrating out the fluctuations, one gets on the basis of symmetry ``an induced gravitational action''~\cite{adler}
\be
e^{iS_{\rm ind}} \equiv \int {\rm D}\varphi\ e^{iS_{\varphi, g_{ab}}} ~,
\ee
where on the grounds of induced diffeomorphism invariance
\be
S_{\rm ind} = \kappa \int \d{{}^4x}\ \sqrt{-g}\ (-2 \Lambda + R(g) + \ldots) ~.
\ee
Here $\kappa$ is the induced (inverse) of the gravitational constant.
This Sakharov-like induced action (and not a Wilsonian effective action) has the usual features (and problems) associated with running of the gravitational and the cosmological constant~\cite{adler}.
Yet, very na\"{\i}vely, this procedure does suggest the existence of effective diffeomorphisms at low-energy.

To conclude, the main point of this section is that from the perspective of the acoustic metric the velocities appear as shifts and that fluctuations of velocities might be related to the fluctuations of the shifts, and thus a general fluctuating geometry or, in other words, in general, a spacetime foam.
In the case of general topology-changing configuration, which defines spacetime foam, shifts, the Lagrange multipliers for the primary momentum constraint, can also fluctuate opening a possibility for a universal scaling of their fluctuations as a function of some characteristic length scale.

Given this picture, the main idea would be to relate the universal geometric properties of spacetime foam to turbulence and discuss issues like Kolmogorov's scaling in the gravitational context.

\section{Wheeler--DeWitt equation and spacetime foam}

The issue of topology change and spacetime foam can be discussed from the canonical and euclidean points of view.
Here we review the relation between the Wheeler--DeWitt equation of the canonical Hamiltonian formalism of quantum gravity and spacetime foam.
We want to discuss the canonical Wheeler--DeWitt equation~\cite{wdw}
\be
H\Psi_\Lambda = 0
\ee
on a spacetime with cosmological constant $\Lambda$.
(In fluid dynamics it seems that the bare value of $\Lambda$ vanishes, but in general this value can be renormalized, together with the gravitational constant in the induced gravity action~\cite{adler}.)
Here $H=0$ is the classical Hamiltonian constraint in the canonical formalism of general relativity.

We begin by writing the spacetime metric in a local neighborhood in the ADM form, which also appears in fluid dynamics:
\be
\d{s}^2 = g_{\mu\nu}\, \d{x^\mu} \d{x^\nu} = N^2 \d{t}^2 - h_{ij} (\d{x^i} + N^i \d{t}) (\d{x^j} + N^j \d{t}) ~.
\ee
Note that in the fluid dynamics context
\be
- N^i \to v^i ~, \qquad N^2 \to c^2 ~.
\ee
In the following, we keep this simple dictionary in mind.
Note that, in accordance with induced diffeomorphisms at long distances, we use the general three-dimensional metric $h_{ij}$ in the expression for the ADM metric.

We find that the extrinsic curvature $K_{ij}$ is
\be
K_{ij} = -\frac{1}{2N} \left( \pa_t h_{ij} + \nabla_i v_j + \nabla_j v_i \right) ~,
\ee
which can obviously be rewritten as the evolution equation
\be
\pa_t h_{ij} = -2N K_{ij} - \nabla_i v_j - \nabla_j v_i ~.
\ee
We also have the Hamiltonian and momentum constraints
\bea
&& H = R^{(3)} + K^2 - K_{ij} K^{ij} - 2\Lambda = 0 ~, \\
&& M_i = \nabla_j K^j_i - \nabla_i K = 0 ~,
\eea
and a second evolution equation
\be
\pa_t K_{ij} = N R^{(3)}_{ij} + N K K_{ij} - 2 N K_{ik} K^k_j - \nabla_i \nabla_j N - \nabla_i v^k K_{kj} - \nabla_j v^k K_{ki} - v^k \nabla_k K_{ij} - N \Lambda h_{ij} ~,
\ee
where $K = h^{ij} K_{ij}$ and $R^{(3)}_{ij}$ and $R^{(3)}$ are the Ricci and scalar curvatures of the spatial metric $h_{ij}$.
The Schr\"odinger equation then is
\be
\frac12 \left( R^{(3)} + K^2 - K_{ij} K^{ij} \right) \Psi_\Lambda = \Lambda \Psi_\Lambda ~.
\ee

It is convenient to rewrite this in a slightly different form.
Following~\cite{wdw} and~\cite{by2}, define
\be
G_{ijkl} = \frac{1}{2\sqrt{h}} (h_{ik} h_{jl} + h_{il} h_{jk} - h_{ij} h_{kl}) ~.
\ee
Define the conjugate momentum to the spatial metric $h_{ij}$ as
\be
\pi_h^{ij} := -i\hbar\frac{1}{\sqrt{h}} \frac{\delta}{\delta h_{ij}} ~.
\ee
We will restore powers of $\hbar$ and put $\kappa = 8\pi G_N = \hbar M_P^{-2}$.
Dimensional analysis tells us that $\pi_h^{ij}$ has units $M L^{-2}$.
The functional Schr\"odinger equation is the Wheeler--DeWitt equation
\be
\left( -2\kappa\hbar^2 \frac{1}{\sqrt{h}} G_{ijkl} \frac{\delta}{\delta h_{ij}} \frac{\delta}{\delta h_{kl}} -\frac{1}{2\kappa} R^{(3)} + \frac{1}{\kappa} \Lambda \right) \Psi_\Lambda[h] = 0 ~.
\ee

Now, in the presence of topology change~\cite{horowitz}, and thus spacetime foam this equation has been argued to become non-linear~\cite{thirdq}.
For example, in the case of a cubic vertex we have the following non-linear Wheeler--DeWitt equation
\be
\left( -2\kappa\hbar^2 \frac{1}{\sqrt{h}} G_{ijkl} \frac{\delta}{\delta h_{ij}}
\frac{\delta}{\delta h_{kl}} -\frac{1}{2\kappa} R^{(3)} + \frac{1}{\kappa} \Lambda \right)
\Psi_\Lambda[h] = c \Psi_\Lambda[h] * \Psi_\Lambda[h] ~.
\ee
where $c$ is an effective coupling constant.
This equation captures the quantum dynamics of spacetime foam in the most direct way.

What would be the meaning of this wave functional in the turbulence context?
One obvious suggestion is that the natural probability density defined by the wave functional,
{\em i.e.}, the probability measure for the spatial three-geometry defined by $h_{ij}$ to be found with spacetime volume $V$ in the region of superspace with volume element $\d\mu[h]$:
\be
\d{P} = |\Psi_\Lambda[h]|^2 \d\mu[h] ~,
\ee
should correspond to the non-Gibbsian stationary probability density in the infinite Reynolds number regime.
Thus sample solutions of the Wheeler--DeWitt equation can provide us with models of non-Gibbsian distribution for turbulence, given the dictionary between fluid dynamics and gravity.

Now, one puzzling feature of this natural proposal is the apparent absence of shifts (velocities) in the expression for the probability density.
This distribution in the case of turbulence should be defined over velocity fields.
In the quantum gravity context the shifts should fluctuate.
This seems to be the case, because they are Lagrange multipliers for primary constraints, which in the case of topology change can become dynamical (see~\cite{morgan}).
We also know that in quantum field theory Lagrange multipliers can become dynamical and acquire vacuum expectation values (a good example is what happens in the large-$N$ $O(N)$ sigma model~\cite{polyakov2}).
This is precisely what we need: correlation functions which depend on the characteristic scale for our Lagrange multipliers, the shifts, which are the fluid velocities.
Finally, the partition function, and thus the wave function becomes in the quantum case a function of the Lagrange multipliers (again, we recall the example of the partition function of the large-$N$ $O(N)$ sigma model).
Similarly, one encounters condensation of Lagrange multipliers in the treatment of string theory as a theory of random surfaces, formulated as a $(1+1)$-dimensional gravity coupled to matter fields (see~\cite{polyakov2}).

Thus we expect that upon the inclusion of topology change
\be
\Psi_\Lambda[h] \to \Psi_\Lambda[h, v^i] ~.
\ee
The shift Lagrange multipliers (the fluid velocities) condense and obtain non-zero vacuum expectation values
\be
\langle v^i(\ell) v^j(0) \rangle \sim \ell^{\alpha} ~,
\ee
where $\ell$ denotes the characteristic scale and $\alpha$ is the critical exponent to be computed from the explicit model of the spacetime foam.
The point of the next section is to argue that holographic models of spacetime foam in $3+1$ dimensions lead to $\alpha = 2/3$, thus reproducing the Kolmogorov scaling.
The non-Gibbsian stationary distribution should then be computed as
\be
\d{P_{v^i}} = |\Psi_\Lambda[h, v^i]|^2\, \d\mu[h] ~.
\ee
In principle such a distribution determines the computation of all correlators.
For example
\be
\langle v^i(\ell) v^j(0) \rangle \equiv \int {\rm D}v^i\ P_{v^i} v^i(\ell) v^j(0) ~.
\ee
Of course, the central question is whether Kolmogorov's scaling follows from this non-Gibbsian distribution.

In the next section we want to argue that even without knowing the explicit form of the distribution, holography constrains the universality of the scaling law in $(3+1)$-dimensional turbulence in accordance with Kolmogorov's theory.

\section{Spacetime foam, holography, and Kolmogorov}

What is known about spacetime foam?
Let us start with the review of~\cite{ng}.
If spacetime is foamy due to quantum fluctuations, the fluctuations $\delta \ell$ will show up when we measure a distance $\ell$, in the form of uncertainties in the measurement.
One way to find $\delta \ell$ is to carry out a gedanken experiment to measure $\ell$~\cite{ng}.
Alternatively we can use a global positioning system to find $\delta \ell$ by mapping out the geometry of spacetime for a spherical volume of radius $\ell$ over the amount of time $T = 2\ell/c$ it takes light to cross the volume.
Let us fill the space with clocks, exchanging signals with other clocks and measuring the signals' times of arrival.
This process of mapping the geometry of spacetime is a kind of computation.
Hence the total number of operations, including the ticks of the clocks and the measurements of signals, is bounded by the Margolus--Levitin theorem in quantum computation~\cite{ml}, which stipulates that the rate of operations for any computer cannot exceed the amount of energy $E$ that is available for computation divided by $\pi \hbar/2$.
A total mass $M$ of clocks then yields, via the Margolus--Levitin theorem, the bound on the total number of operations given by $(2 M c^2 / \pi \hbar) \times 2 \ell/c$.
To avoid black hole formation, in $D$ spacetime dimensions, $M$ must be less than $\ell^{D-3} c^2 /2 G_D$.
Together, these two limits imply that the total number of operations or events that can occur in a spatial volume of radius $\ell$ for a time period $2 \ell/c$ is no greater than $(\ell/\ell_P)^{D-2}$, where $\ell_P \equiv (\hbar G_D/c^3)^{1/(D-2)}$ is the Planck length, and we have dropped multiplicative factors of order one.
In other words, if one regards the elementary events partitioning the spacetime volume into ``cells,'' then the number of cells is bounded by the surface area of the spatial region (corresponding to the holographic scaling of black hole physics~\cite{bek}), and each cell occupies a spacetime volume of $(\ell^D/c) / (\ell/\ell_P)^{D-2} = \ell^2\, \ell_P^{D-2}/c$ on average.
The maximum spatial resolution of the geometry is obtained if each clock ticks only once during the entire time period $\ell/c$.
Then on average each cell occupies a spatial volume no less than $\ell^{D-1} / (\ell / \ell_P)^{D-2} = \ell\, \ell_P^{D-2}$, yielding an average separation between neighboring cells no less than $(\ell\, \ell_P^{D-2})^{1/(D-1)}$.
This spatial separation is interpreted as the average minimum fluctuation of a distance $\ell$.

One of the points of~\cite{ng} is that in the case where space and time are treated on different footing (this is natural from the point of view of turbulence).\footnote{
Also, from the point of view of quantum gravity, this is natural in a dynamical regime of emergent spacetime.}
The scaling of length in the simple holographic models of spacetime foam~\cite{ng} is as follows:
\be
\delta \ell \sim \ell^{1/(D-1)}\, \ell_P^{(D-2)/(D-1)} ~.
\ee
Note that it is natural to expect that the coefficient multiplying $\ell\, \ell_P^{D-2}$ for $\langle \delta\ell^{D-1} \rangle$ is universal, being given by the holographic principle.
In $D=3+1$ dimensions, consider a cube of size $\ell \times \ell \times \ell$.
The number of degrees of freedom that the cube can contain is given by $\ell^3 / \delta \ell^3$, which is bounded by the requirement that the entropy $S/k_B \sim (1/4) (6 \ell^2)/ \ell_P^2$, where $6 \ell^2$ is the surface area of the cube.
(In fact this is one way to get $\delta \ell \sim \ell^{1/3} \ell_P^{2/3}$.)
Unfortunately there is a small ambiguity in the determination of the coefficient.
The ambiguity comes about because it is not clear whether one should use a big cube (volume $\ell^3$) containing the small cubes (volume $\delta \ell^3$) to do the counting of the degrees of freedom or a big sphere (volume $4 \pi\ell^3/3$) containing small spheres (volume $4 \pi\delta \ell^3/3$).
The holographic principle argument uses spheres, but the packing of small spheres in a big sphere is not tight (having space between neighboring small spheres).
But in any case, the coefficient is positive.
The upshot is that provided one defines the velocity as
\be
v \sim \frac{\delta \ell}{t_c} ~,
\ee
where the natural characteristic time scale is
\be
t_c \sim \frac{\ell_P}{c} ~.
\ee
It follows that
\be
v \sim c \big(\frac{\ell}{\ell_P}\big)^{1/3} ~.
\ee
Then it is obvious that a Kolmogorov-like scaling has been obtained, {\em i.e.}, the velocity scales as $v \sim \ell^{1/3}$ and the two-point function has the needed two-thirds power law.\footnote{
The energy dissipation rate $\varepsilon$ is nothing but $c^3/\ell_P$, where now $c$ is effective (it is the speed of sound) and $\ell_P$ is effective (it is given by the induced gravitational constant).}

The other consequences of this scaling discussed in Section~2, also known from Kolmogorov's work, would follow.
Note that the relation between turbulence and gravity, as discussed so far, is in the same number of dimensions.
Yet the full gravitational dynamics is only induced at long distance, and in principle is ill-defined at short distance.
As opposed to the usual prescription of Wilsonian effective field theory where we systematically integrate out those degrees of freedom, the high-energy limit seems particularly natural here because in the map between turbulence and spacetime foam, the foam is an UV concept.
Thus, we should expect the inverted RG scaling noted in~\cite{gaw}.

Now, if the connection between turbulence and spacetime foam can be indeed established as indicated above, then the emergence of a Kolmogorov-like scaling might not be simply fortuitous.
Yet, we note that this observation has been made for three spatial dimensions.
On the other hand, Kolmogorov's scaling seems to be dimension-independent.
We now contrast the scaling laws for turbulence in $3+1$ and $2+1$ dimensions.

\subsection{$3+1$ vs.\ $2+1$ and the energy cascade}

How does the holographic model of spacetime foam compare to what is known about the r\^ole of the dimensionality of space in turbulent flows, {\em i.e.}, $(2+1)$-dimensional versus $(3+1)$-dimensional scaling laws.
First, in three spatial dimensions, the two-thirds law has been well-tested experimentally (see~\cite{frisch}).
Also, this leads to the four-fifths law for the three point function and what is more important to the following scaling for energy as function of momenta:
\be
E(k) \sim k^{-5/3} ~.
\ee
Note that in general~\cite{frisch} the energy scaling
\be
E(k) \sim k^{-n}
\ee
is related via a {\it one-dimensional} Fourier transform to the scaling of the two-point function for the velocity field
\be
\langle (\delta v(\ell))^2 \rangle \sim \ell^{n-1} ~.
\ee
Thus the two-thirds scaling of the two-point function leads to the $k^{-5/3}$ scaling in momentum space.
As shown above, applying the holographic principle to enumerate the degrees of freedom of spacetime foam does reproduce the Kolmogorov scaling.

In $2+1$ dimensions, Kraichnan~\cite{frisch, bob} has argued that the relevant scaling law associated with the energy cascade is also the $k^{-5/3}$ Kolmogorov law.
But, as Kraichnan crucially observed, the energy cascade is inverted in $2+1$ dimensions as opposed to $3+1$.
Related to this is the fact that in $2+1$ dimensions, there is another conserved quantity apart from energy,
{\em i.e.}, enstrophy~\cite{bob1}
\be
\Omega = \int \d{{}^2}x\ \omega^2 ~,
\ee
where $\vec{\omega} \equiv \nabla \times \vec{v}$ is the vorticity.
By repeating the Kolmogorov like reasoning for the enstrophy Kraichnan~\cite{bob1, bob} obtained $v \sim l$ which leads to the $k^{-3}$ scaling in momentum space.

How does this compare with holographic spacetime foam?
In the $D=3+1$ case, we seem to have an agreement with Kolmogorov's scaling.
But in the $D=2+1$ case, the na\"{\i}ve holographic model of spacetime foam gives
\be
v \sim c \big(\frac{\ell}{\ell_P}\big)^{1/2} ~,
\ee
and thus $v^2 \sim \ell$, which in momentum space implies the $k^{-2}$ scaling instead of Kolmogorov--Kraichnan's $k^{-5/3}$.\footnote{
The prediction of holography in $2+1$ dimensions, that the scaling of the energy in momentum space is $k^{-2}$, might have to do with the UV completion of the holographic model.
Provided that we understand M-theory beyond the eleven-dimensional supergravity limit, but as a true quantum theory, the $\ell^{1/3} \ell_P^{2/3}$ behavior of the two-point function might appear naturally and would then be dimension independent.}
Admittedly, the holographic scaling is close to Kraichnan's, which has been well established in numerical simulations~\cite{frisch}.
Nevertheless, after vortices kick in this scaling, due to the conservation of enstrophy~\cite{frisch, bob}, this should change to the $k^{-3}$ scaling.
Na\"{\i}vely there does not seem to exist an obvious analog of this fact on the holographic spacetime foam side, unless one appeals to the topological nature of $2+1$ gravity.

The axial symmetry in $2+1$ dimensions makes it obvious that the vorticity and any power of it are conserved.
That is to say that $v \sim \ell$ by dimensions of the vorticity, and then $t \sim {\rm constant}$ if one defines $v \sim \ell/t$.
As $2+1$ gravity is topological, there is no time evolution, so constant $t$ is indeed the expected scaling.
Thus the conservation of enstrophy is equivalent to the topological character of gravity and leads to the $k^{-3}$ power law for the energy spectrum.

To summarize: the na\"{\i}ve holographic spacetime model does not seem quite to match Kolmogorov--Kraichnan's scaling in $2+1$ dimensions.
How about the inverse energy cascade of Kraichnan~\cite{frisch, bob, bob1} in $2+1$?
The only obvious difference between $2+1$ and $3+1$ gravity is that the $2+1$ gravity is topological and that in that case there exists a holographic anomaly~\cite{skenderis}.
Still, it is not clear how to relate these unique features to the inverse cascade.
One possibility is offered by the work of Polyakov on conformal turbulence~\cite{polyakov} which might be holographically dual to a $(2+1)$-dimensional gravitational description.
In Polyakov's discussion the r\^ole of the enstrophy cascade was clearly identified.
Note also that in the case of an AdS/CFT-like holographic map~\cite{malda}, the RG scalings from the bulk of spacetime and the holographic boundary are inverted.
In other words, the UV of the holographic boundary corresponds to the IR of the bulk~\cite{uvir}.
This might offer a way of understanding the inverted cascade in $2+1$ dimensions, provided there indeed exists a $(2+1)$-dimensional AdS-like dual to the conformal turbulence in two dimensions.
In any case, the na\"{\i}ve holographic spacetime model in $2+1$ dimensions has to be modified to take these important physics considerations into account.
On the fluids side, contrary to first appearance, two-dimensional turbulence turns out to be much more complex and richer in physics than the three-dimensional case.
There are several types of cascades at work with interplay between statistics and ({\em e.g.}, coherent) structures~\cite{tabeling}.

One might wonder what the geometric analog of the Kolmogorov's three-point function is in the $(3+1)$-dimensional holographic spacetime foam model and also where the factor $-\frac{4}{5}$ comes from in this model.
Given the picture offered in this note, to figure out the four-fifths law we would need to know the vertex for the spacetime foam, which is beyond the simple scaling relations implied by holography.

Finally, we emphasize that the holographic scaling is semiclassical.
Loop corrections might naturally correspond to turning on the viscosity.
From this point of view, viscosity is a loop expansion parameter (an effective $\hbar$).
These dynamics would indeed be interesting to consider.

\section{Future directions and conclusions}

In this note we have argued, based on the analog fluid models of gravity, that in the gravitational context the phenomenon of turbulence is intimately related to the properties of spacetime foam.
In particular, using some general observations about holographic models of spacetime foam we have discussed the relation between turbulence and holography and the interpretation of the Kolmogorov scaling in the quantum gravitational setting.\footnote{
Note that this dictionary is natural from the point of view of the proposed general relation between quantum gravity and non-equilibrium statistical physics~\cite{timem}.}

The duality between fluids and spacetime foam discussed in this note exists in the same number of dimensions.
One might wonder whether turbulence is dual to a classical gravitational background following the philosophy of the AdS/CFT correspondence.
Two ideas come to mind.

First, as we have previously mentioned, Polyakov has considered a two-dimensional CFT in the context of $(2+1)$-dimensional turbulence~\cite{polyakov}.
Searching for a gravitational AdS-like dual of this two-dimensional CFT seems natural from the point of view of this paper.
The natural gravitational dual, according to our proposal, should involve spacetime foam.
In this respect we note recent papers on topology and AdS/CFT~\cite{adsworm}.
One upshot is these investigations is that wormhole configurations can be accounted by the correspondence and are not of the third quantization type.
From the point of view advocated in this note, these wormhole configurations might be used as models of spacetime foam in the AdS/CFT context, and should provide dual gravitational backgrounds for turbulent fluid dynamics on the boundary.

Secondly, spacetime foam has also been discussed in string theory in the context of the microstate picture of black holes.
(For reviews, see~\cite{mathur, jan} and references therein.)
The entropy of a black hole is determined by the area of the event horizon: $S_{\rm BH} = A/4G_D\hbar$~\cite{bek}.
The thermodynamic description of a black hole originates in an underlying theory of gravitational statistical mechanics:
there are $e^{S_{\rm BH}}$ microstates that one associates to a black hole.
Models for this may be considered in the context of the AdS/CFT correspondence.
For example, the physics of half-BPS black holes with ${\rm AdS}_5\times S^5$ asymptotics is described by operators in the dual ${\cal N}=4$ $SU(N)$ super-Yang--Mills gauge theory that preserve sixteen supercharges~\cite{llm}.
In terms of these boundary data, there is a density function on the phase space of the gauge theory such that when we integrate against this kernel $\rho$, we reproduce bulk correlators in the semiclassical limit:
\be
\lim_{\stackrel{N\to\infty}{\hbar\to 0}} ({\rm tr}(\rho {\cal O}) - \vev{{\cal O}})\to 0 ~,
\ee
for macroscopic observables ${\cal O}$.
Similar behavior applies to higher point correlators.
At $\hbar$ precision, the correlation functions of a probe operator in the thermal density matrix equates to the correlation functions in a pure state that is an element of the thermodynamic ensemble for almost all probes.\footnote{
This statement is made precise for the $M=0$ BTZ black hole in~\cite{bks}.} 
Thus, the entropy measures the inability to distinguish elements of the ensemble from each other.
The semiclassical geometry emerges from a thermodynamic {\em coarse-graining} over the microstates~\cite{bdjs}.
In particular, a new scale in quantum gravity associates to the semiclassical horizon~\cite{lm}.
The ``typical'' state corresponds to a spacetime foam, with topologically complex features at the scale of the horizon;
the effective long wavelength description in low-energy gravity is a singular geometry~\cite{bdjs}.
Once again, from our point of view, this ``typical'' spacetime foam state should be dual to a boundary turbulent flow.
More precisely, via the AdS/CFT dictionary, the generating functional of correlators in the infinite Reynolds number limit would be determined by the semiclassical form of the wave functional for the ``typical'' spacetime foam state.

Finally, non-Abelian hydrodynamics has been treated in a dual AdS-like way in the recent literature as mentioned in the beginning of this note~\cite{sonetal, shiraz, makoto}.
(See also an illuminating review~\cite{sson}.)
In this context, the hydrodynamic description is related to black hole backgrounds.
Indeed, in~\cite{vr} the technology of~\cite{shiraz} is adapted to apply $(2+1)$-dimensional fluid dynamics to yield a long-wavelength description of black holes in ${\rm AdS}_4$.
In view of the above-mentioned microstate picture of black holes we should expect that non-Abelian hydrodynamic turbulence should be dual to the ``typical'' spacetime foam state.

Obviously there are many future avenues for working out the proposal presented in this note.
Through the above mapping between turbulence and spacetime foam and its possible further elaborations, we hope that turbulence, the great problem of classical physics may be informed by quantum gravity, the great problem of quantum physics, and of course, vice versa.

\section*{Acknowledgments}

We thank Esko Keski-Vakkuri, Tatsu Takeuchi, and Uwe Tauber for discussions of the ideas presented in this note.
VJ is grateful to the LPTHE, Jussieu and the Helsinki Institute of Physics for generous hospitality.
DM is supported in part by the U.S.\ Department of Energy under contract DE-FG05-92ER40677.
YJN is supported in part by the U.S.\ Department of Energy under contract DE-FG02-06ER41418.

\end{document}